\documentstyle[11pt,newpasp,twoside,epsf]{article}

\def\edcomment#1{\iffalse\marginpar{\raggedright\sl#1\/}\else\relax\fi}
\reversemarginpar

\newcommand{\kms}{\hbox{\,km\,s$^{-1}$}}

\begin{document}

\title{On the X-ray emission of $\eta$ Carinae's outer ejecta}
{\author{Kerstin Weis\footnotemark}}
\affil{University of Minnesota, Astronomy Department, 116 Church
Street SE, Minneapolis, MN 55455, USA}

{\footnotetext{Feodor-Lynen fellow of the Alexander-von-Humboldt foundation}}

\author{Michael F. Corcoran}
\affil{High Energy Astrophysics Science Archive Research Center,
Goddard Space Flight Center, Greenbelt, MD 20771, USA}

\author{Kris Davidson}
\affil{University of Minnesota, Astronomy Department, 116 Church
Street SE, Minneapolis, MN 55455, USA}

\begin{abstract}
The extremely luminous and unstable star $\eta$ Carinae is 
surrounded by ejecta formed during the star's giant 
eruption around 1843. The optical nebula consists of an inner 
region, the bipolar Homunculus and the outer ejecta. 
The X-ray emission as detected in ROSAT and CHANDRA 
shows a hook shaped emission structure mainly at the position of the
outer ejecta substantially larger than the Homunculus. 
We present results of a comparative study of 
the optical morphology, the kinematics and the X-ray emission 
of the outer ejecta around $\eta$ Carinae. In general we find that
the X-ray emission traces the shocks of faster moving knots in the
outer ejecta.
First results of CHANDRA/ACIS data will be presented, with 
spectra of selected areas in the outer ejecta giving insight to the 
conditions (temperature, present elements and degree of ionization)
of the hot gas. The X-ray spectra will again be compared to the kinematics 
of the gas as known from our optical spectra.
\end{abstract}

\section{Introduction}

Among the brightest and most likely also most massive stars, 
$\eta$ Carinae is an outstanding object. For a very
detailed description on various aspects of this star the reader is referred to
several reviews, especially Davidson \& Humphreys (1997) and references 
therein. Recently, numerous observations in the optical and X-ray band
gave hint to a possible binary system in this object 
(Damineli 1996, Damineli et al.\ 1997, Davidson et al.\ 2000, 
Corcoran et al.\ 2001), but since the nature of the  central 
object will be of no further interest to this work, 
it will therefore be ignored in the following. 
Whether $\eta$ Car is a binary or not, at least one of the 
components of the system had a mass above 50\,M$_{\sun}$ and evolved
quickly into a supergiant and encountered an unstable phase with 
high mass loss as an LBV type star. In 1843 the star underwent a 
so-called giant eruption in which a large amount of mass was 
peeled off. The strong stellar wind together with the giant eruption 
led to the formation of a circumstellar nebula first detected around
1950 (Gaviola 1946, 1950 and Thackeray 1949, 1950) 
Due to the nebula's appearance of a little man Gaviola called 
it accordingly the {\it Homunculus}. The nebula is now larger 
and is separated into two different parts, the inner
bipolar nebula still called the Homunculus and the very filamentary
{\it outer ejecta}. The outer ejecta---the section of major interest in
this study---represent a large collection of bullets, knots
and filaments distributed out to a distance of 30\arcsec\ in radius 
around $\eta$ Car.
Both sections of the nebula are expanding, the Homunculus with about 650\,\kms\
(Davidson \& Humphreys 1997 and references therein, Davidson et al.\ 2001), 
the outer ejecta with velocities ranging from a few \kms\ to more than 
2000\,\kms\,(Meaburn et al.\ 1987, Meaburn et al.\ 1996,  Weis 2001a, 
Weis 2001b, Weis et al.\ 2001).

\begin{figure}
\plottwo{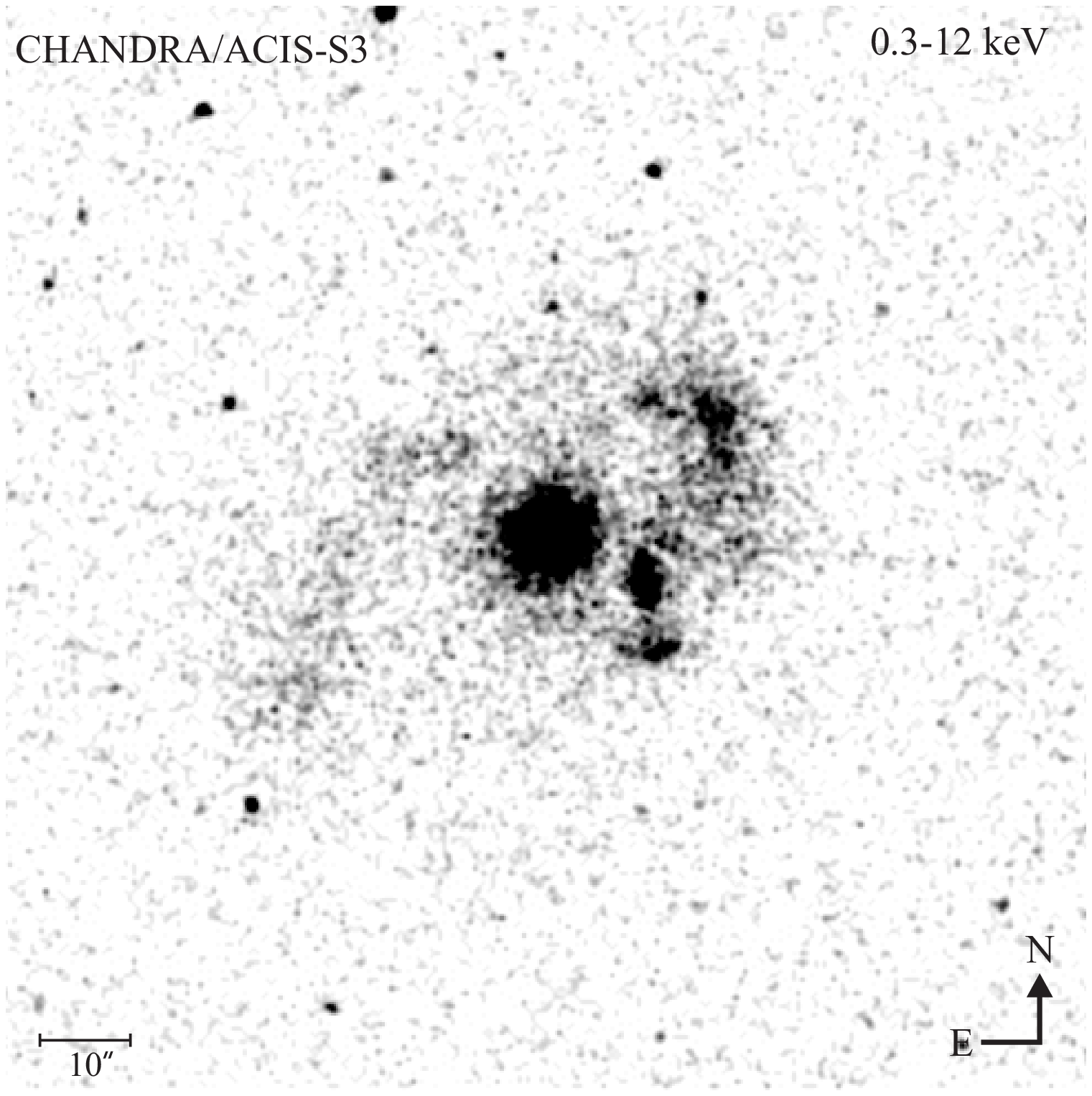}{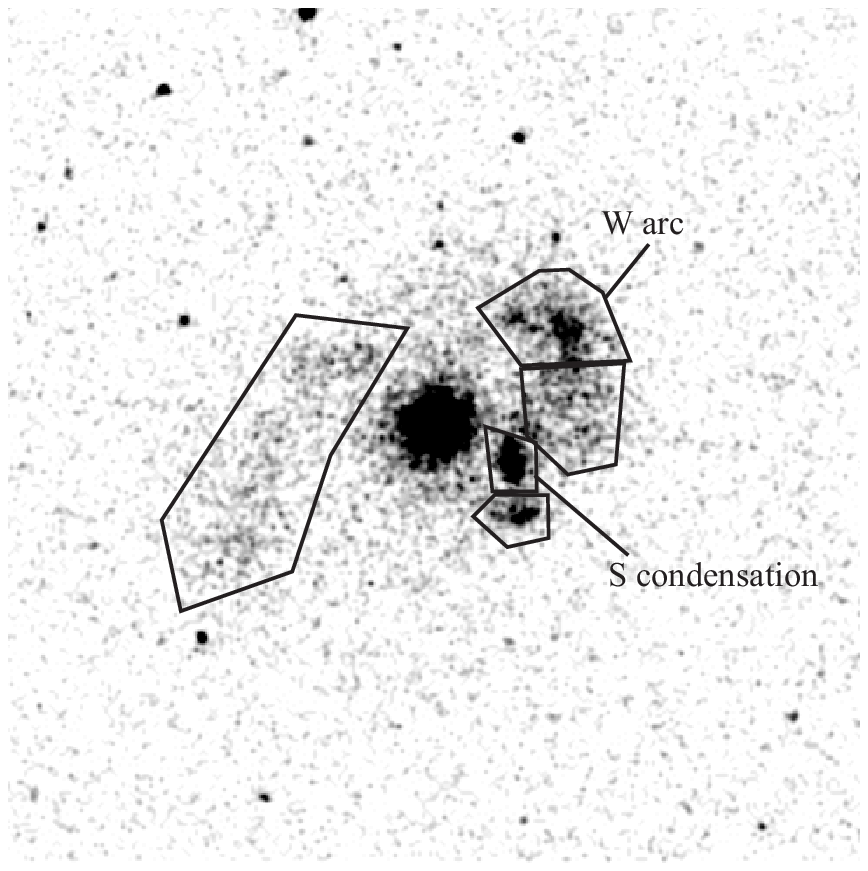}
\caption{The figure (left) shows a CHANDRA image 
between 0.3 and 12 keV of the outer
  ejecta and the central source and the same image (right) with the regions
  indicated selected to extract spectra. Regions marked and named are
  discussed in more details in the text.
}
\end{figure}

The existence of X-ray emission of the $\eta$ Car complex was known 
since early observations with a proportional counter system aboard a {\it
Terrier-Sandhawk} rocket (Hill 1972). Later with the
{\it Einstein Observatory} the resolution of X-ray telescopes  improved
to a few arcseconds and made it possible to spatially distinguish 
several sources. $\eta$ Carinae was detected as an extended X-ray
source (Chlebowski et al.\ 1984) in the soft and hard bands.
With the launch of the {\it R\"ontgensatellit\/} (ROSAT), resolution and 
sensitivity became even
better in the soft X-ray regime and it could be proven now that the harder
and softer X-ray component of $\eta$ Car and its nebula (c.f.,
Corcoran et al.\ 1994, 1995, 1996, 1997), are spatially
distinct: a softer diffuse shell of the nebula and a harder
point-like source centered on $\eta$ Car. The hard  X-ray central
source also  shows variability and a point-like
character (Corcoran et al.\ 1995). In comparing the X-ray
emission detected by ROSAT and the optical emission as seen in the HST
images with the kinematics of the outer ejecta it was recently shown
that a correlation exists between the X-ray emission and the expansion
velocities (details see Weis et al.\ 2001). 
X-ray emission is brightest at places where 
high velocities are seen. The unusual 
morphology of the X-ray emission (hook shaped compared to the bipolar
Homunculus and irregular and filamentary outer ejecta) can therefore
be explained by the X-ray emission in shocks produced by
the faster moving knots in the outer ejecta (Weis et al.\ 2001). 
The X-ray emission traces the shocks of the expanding outer ejecta.

\section{Observations}  

Here we will report on the first results concerning the outer ejecta
of $\eta$ Carinae observed with CHANDRA's {\it Advanced CCD Imaging
Spectrometer\/} (ACIS) using the 0th order image of the {\it High Energy
Transmission Grating\/} (HETG) on the S3 chip. The spatial
resolution is (unbinned) 0\farcs51 
and the exposure time of our observations (PI:MFC) was 100 ksec.  
Using the 0th order image we made use of the fact that this ACIS image also 
provided us with not only spatial but also spectral resolution. Therefore we
extracted spectra from selected regions of the outer ejecta. Of the 5 regions
originally selected spectra, results from two regions will be 
presented here: The {\it S\,condensation} and the {\it W\,arc}. 
The regions are 
named based on existing naming conventions by Walborn (1976). 
The positions of the regions, and their measured areas can be seen in 
Fig.\ 1 
(right image). Note that both regions roughly coincide with the two brightest 
X-ray knots detected with ROSAT (Weis et al.\ 2001). 

\begin{figure}
\plottwo{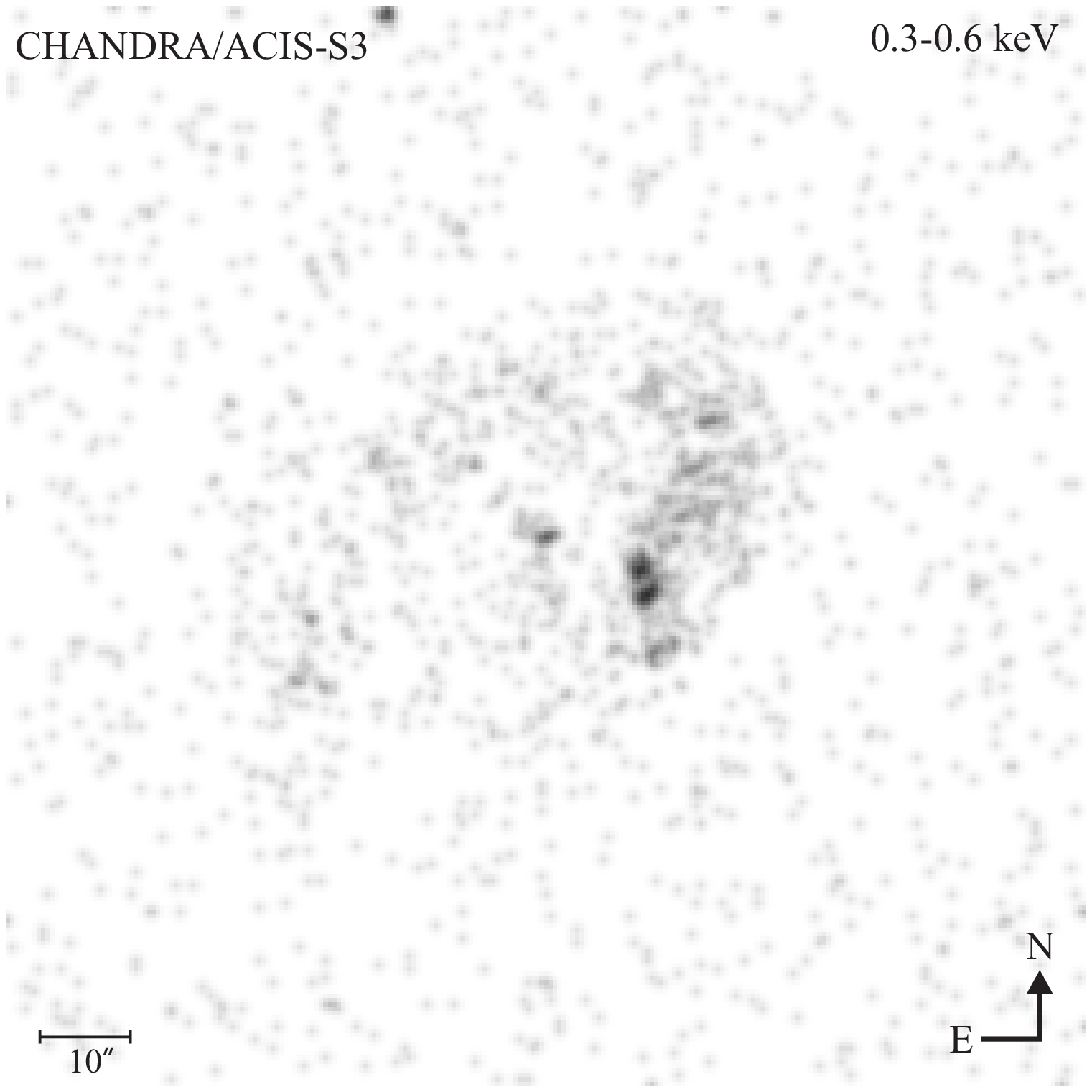}{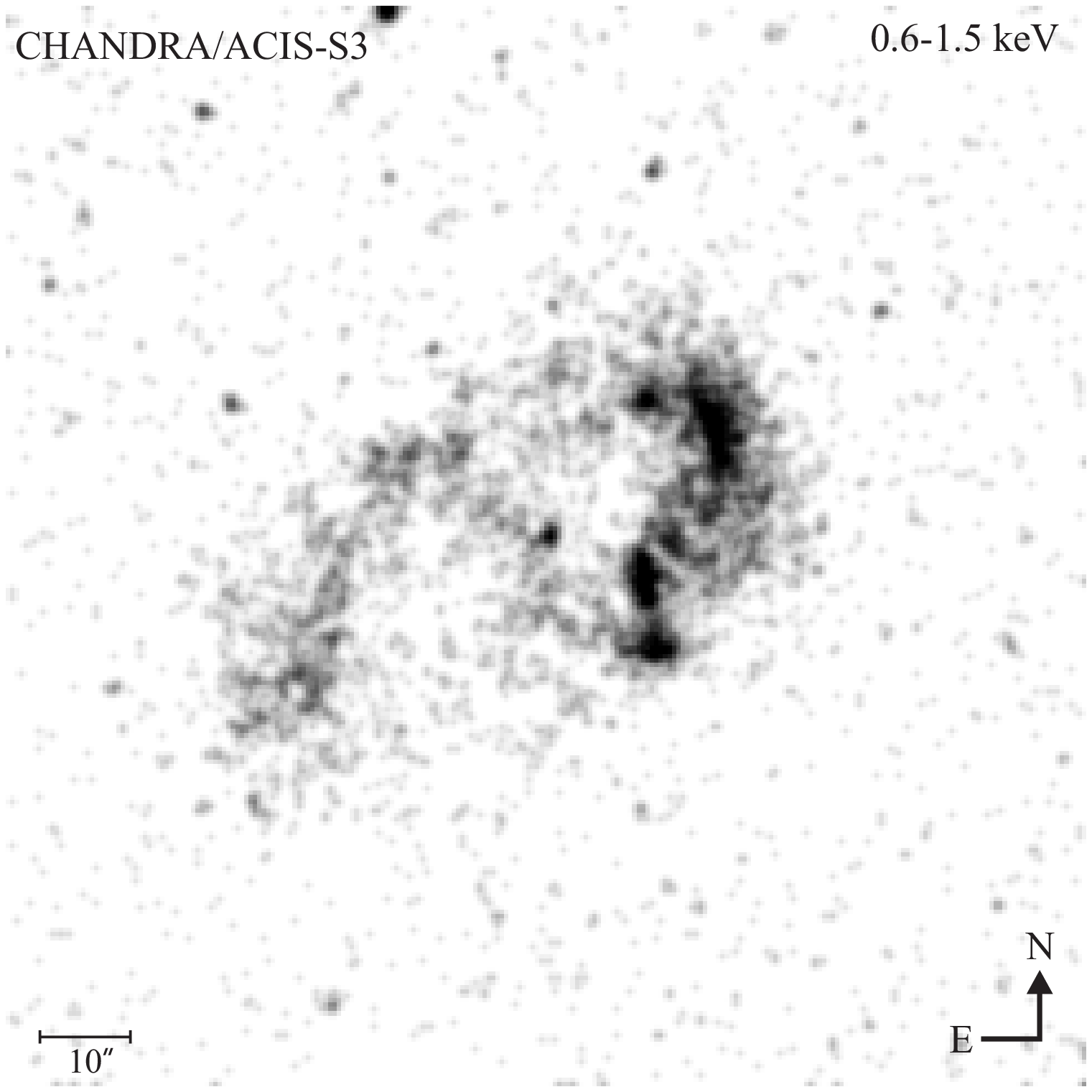}
\caption{CHANDRA X-ray images now in the 0.3-0.6 keV (nitrogen-line-image, 
see text) and the 0.6-1.5 keV bands. 
}
\end{figure}

\section{First results and their implications}

Fig.\ 1 (left) shows a CHANDRA image
of $\eta$ Carinae. The emission detected here ranges from 0.3 to 12 keV and
shows the characteristic hook shaped structure of the outer nebula. Within
the hook a point-like the central source is visible, which might be 
marginally extended. With the high-resolution of CHANDRA compared to 
ROSAT the originally
identified knots 1 and 2 (see Weis et al.\ 2001) can be subdivided into 
more knots and brighter regions. The brightest regions, beside the central
source, is the S\,condensation which itself seems two consist of two 
separated maxima.
This becomes especially obvious in the left panel in Fig.\ 2.
In this figure the X-ray emission in two different energy bands is displayed.
The left panel shows the emission between 0.3-0.6 keV. 
The energy band is dominated by the hydrogen-like nitrogen line at 0.51 keV,
except in the hot central source. In each extracted spectrum this line
is isolated and well separated from the brighter emission at higher energies
(see example spectra in Fig.\ 3).
Most of the diffuse thermal emission occurs at energies between 0.6 and
1.5 keV, shown in the right panel of the figure.  Fig.\ 2
shows some differences between the two energy ranges;
in particular, the nitrogen emission is strong in the 
S condensation southwest of the star.  This is consistent with
optical and UV measurements (Davidson et al.\ 1986, Dufour 1989). The right 
panel shows not only the hook-shaped structure of the X-ray nebula, but
also a zig-zagged bridge-like structure near the center.
This seems to extend from the northeastern part of the
hook, through the central source, to the southwestern part 
(the S condensation).

\begin{figure}\plotone{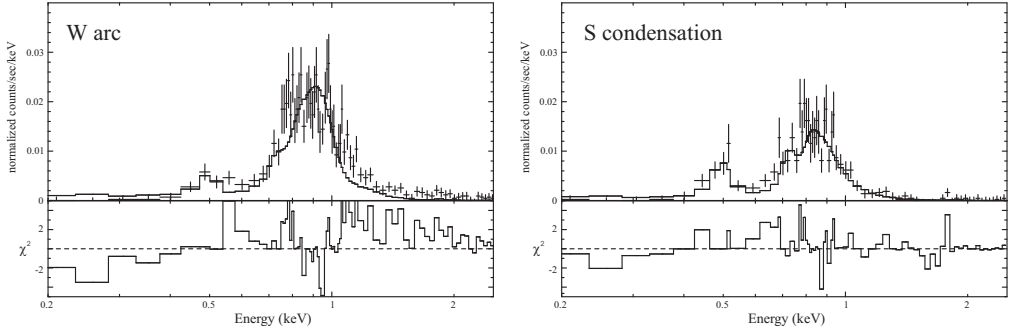}
\caption{This figure shows the CHANDRA/ACIS  X-ray spectra as of the 0th order
  of the HETG of two selected regions which were marked in
  Fig.\ 1} 
\end{figure}

For the S\,condensation and the W\,arc region spectra 
have been extracted and are shown 
in Fig.\ 3. Both spectra are background subtracted and 
were fitted using a Mewe-Kaastra-Liedahl thermal plasma model 
 within the XSPEC task. For both
spectra a hydrogen column density of $\log$\,N$_{\rm H}$\,$= 21.3$ was 
assumed. For all elements but nitrogen and iron, solar abundances 
were taken. The former elements were left open to fit. 
A higher nitrogen abundance is
expected in the nebula due to the CNO processed material in the nebula around
$\eta$ Carinae. Even though some sections and most likely some lines in 
the X-ray spectra were not fitted, the overall fit with this equilibrium 
model is in agreement with the data. For the W\,arc we therefore derived 
a temperature of kT=0.78 and a nitrogen overabundance of a factor of 40. For 
the S\,condensation the temperature is slightly lower, and also more 
typical for the ejecta, with kT=0.64 while the nitrogen abundance is 
enhanced by an unbelievable factor of 118.
We also checked our data fits using a Raymond-Smith
model, but the agreement with the data where worse. 
All other spectra extracted showed similar results: a temperature around 0.63
keV and a nitrogen enhancement of a factor of 20-40. 
The higher temperature in the
W\,arc is consistent with the fact that we found the highest expansion
velocities---which reach up to 2000\,\kms---in this area. However, the 
highest temperatures derived from the spectra, around 0.8 keV, are  lower 
than what would be
expected from the kinematics, which lead to post-shock plasma temperature 
of 1.5 to 2 keV. Even though such high temperature gas has not been detected
so far, additional emission around 1.1 keV in the spectra of the W
Arc is visible and was not fitted by the model. At least part of this emission
might be attributed to hotter gas from the faster knots we detect. 
In addition most likely
non-equilibrium effects might play a role (see discussion in Weis et
al.\ 2001).

In summary, we can report the following results: The outer ejecta of 
$\eta$ Carinae show a hook-shaped X-ray emission with more
structure than known before. Previously detected knots can be
identified with the S\,condensation and W\,arc mentioned in earlier discussions
of optical images. 
The X-ray gas, like the cooler matrial seen in  optical spectra,
is nitrogen-rich. Nitrogen X-ray emission is particularly strong
in the S\,condensation. 
The hottest diffuse X-rays appear in the W\,arc, which is also the location
of the highest velocities observed near $\eta$  Car.

\end{document}